\documentclass[a4paper,10pt]{article}

\usepackage[T1]{fontenc}
\usepackage[latin1]{inputenc}
\usepackage{amsmath}
\usepackage{amsfonts}% to get the \mathbb alphabet
\usepackage{amsthm}
\usepackage[english]{babel} 
\usepackage{graphicx}

\newcommand{\C}{\mathbb{C}}
\newcommand{\R}{\mathbb{R}}
\newcommand{\Z}{\mathbb{Z}}

\graphicspath{{./Figures_1003/}}

\title{Analysis and separation of time-frequency components in signals with chaotic behavior}
\author{B. Ricaud*, F. Briolle*, F. Clairet**\\\small{* Centre de Physique Th\'eorique - UMR 6207, Luminy Case 907, F-13288 Marseille Cedex 9 France}\\\small{** Institut de Recherche sur la Fusion Magn\'etique, CEA Cadarache, F-13108 Saint Paul Lez Durance, France}\\\small{email: ricaud@cpt.univ-mrs.fr}}

\begin{document}

\maketitle
\begin{abstract}
The analysis of chaotic signals with time-frequency methods is considered. For this purpose, two new transformations are presented which consist in the decomposition of a signal onto an orthogonal set of respectively linear and hyperbolic chirps. The linear chirp transformation is able to discriminate and extract particular chaotic components in non-stationary square integrable signals. This is demonstrated in an example studying the reflectometry measures of a turbulent plasma. The hyperbolic chirp transformation is designed for the detection and extraction of chaotic parts in self-similar processes such as stochastic motions. Mathematical connections are made between these two methods and other well-known transformations.
\end{abstract}

{\bf
For a rapid detection of chaotic patterns in non-stationary signals, new analysis are necessary. This is what our paper addresses. The links between our new methods and other well-known signal processing transformations are described. Some results on experimental data confirm the adequacy of our methods.
}
\section{Introduction}

Many numerical techniques have been designed to analyze the dynamics of chaotic systems. Among them are the computation of the Poincar\'e section, the estimation of asymptotic quantities such as Lyapunov exponents, entropy or fractal dimension. Although they can give accurate information on the average and long-time behavior of chaotic systems, they may be unable to retrieve particular crucial information. 
In the context of control of chaos for instance, the procedure may involve real-time estimation of the system state and fast detection of chaotic patterns.
In such cases, a signal processing approach with time-frequency methods may be an appropriate choice. For example, the wavelet transform and the scalogram have been extensively used~\cite{Far} to study turbulent signals. The analysis of ridges in the time-scale or time-frequency plane allows to characterize particular chaotic behaviors. Resonance transition or trapping can be detected as well as signs of weak or strong chaos~\cite{Chandre}.

Our work has two goals. The first one is to present a new transformation allowing a fast and efficient detection and extraction of ridges in the time-frequency domain. 
%In this article, we present two transformations able to isolate and extract chaotic parts of a signal. It is based on the extraction of the ridges in the time frequency plane.
It is called the Linear Chirp Transformation (LCT). A key advantage of the method is that a tuning parameter is associated to the transformation. Unlike usual time-frequency methods, it allows to adapt the transformation to the shape of the chaotic components which are to be detected and extracted. The second advantage is its fast computation as it is related to the fast Fourier transform. The inverse transformation, which provides the temporal expression of the extracted ridges, can also be implemented with a fast algorithm.
This transformation is illustrated with an application to the analysis of data issued from the reflectometer of the toroidal fusion machine Tore Supra~\cite{Clairet}. The actual importance of the transformation and the main motivation for its development originate from the need in fusion experiment to obtain an accurate plasma density profile. The edge turbulence must be measured precisely and controlled as it can break the confinement and create plasma leaks. Reflectometry appears to be particularly suited for retrieving this information. It allows to probe the state of the plasma in a fast and accurate way. Nevertheless, the reconstruction of the density profile can be delicate: chaotic transport creates multi reflections in the signal, difficult to interpretate. Recent studies~\cite{BLMV1,BLV2} revealed the efficiency of a chirp-like transformation to filter out undesired echos in such signals and our work presents the improved results obtained with the LCT.

The second goal is to extend the time-frequency analysis to other types of signals. Indeed, the standard tools are dedicated to square integrable signals usually fluctuating around a mean value. But no equivalent method exists for the investigation of stochastic processes such as Brownian or fractional Brownian motions. A connection has been made between self-similar and stationary signals in~\cite{FBA,BAF}: we will show that this opens the door to an application of time-frequency analysis to such processes. We present a second new transformation, called the Hyperbolic Chirp Transformation (HCT). We establishes its connection with the LCT using the results of the previously cited articles. It hence suggests that the HCT can extract chaotic parts in signals issued from stochastic processes in the same manner as the LCT does for square integrable signals.

The definition and properties of the chirp transformations are given in Section~\ref{transform}. The connection between these two transformations is discussed together with their relations to other types of time-frequency methods such as the fractional Fourier, the Mellin or the Lamperti transformation.
Then section~\ref{experiment} exhibits the procedure for the application of the linear chirp transformation along with the different results obtained on an example: the extraction of the plasma reflections in chaotic reflectometry signals. Eventually, in section~\ref{conclusion} some important characteristics of this signal processing technique are discussed.

%%%%%%%%%%%%%%%%%%%%%%%%%%%%%%%%%%%%%%%%%%%%%%%%
\section{The chirp transformations}\label{transform}

In this section we define the new transformations. We derive some of their properties which are important for their use in applications and for efficient numerical implementation. We also connect them to more familiar transformations.

A square integrable function $f$ observed from a different point of view can reveal interesting properties. A convenient way may be to use the projection of the function onto the set $\{\psi_k\}_k$ of (generalized) eigenvectors of a selfadjoint operator. The operator is chosen depending on which signal properties we want to emphasize. For example, the decomposition on the eigenvectors of the operator generator of translation will highlight stationary signals whereas the decomposition on the eigenvectors of the generator of dilatations will underline self-similar signals. 
This has been the point of view adopted in~\cite{MV,MMV} to define the tomogram transforms and it is at the origin of the transformations which are the focus of this paper.
Let $f$ be an element of the Hilbert space $L^2(T_1,T_2)$ ($T_1,T_2$ may be infinity). Its projection on the vector $\psi_k$ is given by the scalar product
\begin{equation}
(f,\psi_k)=\int_{T_1}^{T_2}f(t)\overline{\psi_k(t)}dt.
\end{equation}
Then the spectral theorem states that the signal $f$ can be written as the sum:
\begin{align}\label{resynt}
f(t)=\sum_k(f,\psi_k)\psi_k,
\end{align}
if $k$ is discrete or with an integral if $k$ is continuous. Two important reasons for the choice of the operator approach are:
\begin{itemize}
\item[i)] the equation~\eqref{resynt} provides the inverse transformation associated to the projection. Hence a signal can be reconstructed in a straightforward manner thanks to the orthogonality property of the eigenvectors.
\item[ii)] when the selfadjoint operator depends on a parameter $\theta$, so does its eigenvectors but changing $\theta$ does not affect the orthogonality property. Hence $\theta$ can be used as a tuning parameter. 
\end{itemize}

%%%%%%%%%%%
\subsection{The Linear Chirp Transformation}
\subsubsection*{Definition}
We define the linear chirp transform $C_L^{\theta}f$ of a function $f\in L^2(\R)$ by the scalar product:
$$(C_L^{\theta}f)(x)=(f,\psi_x^{\theta}),
$$
where the $\{\psi_x^{\theta}\}_x$ are orthonormal and have the following expression for $\theta\neq n\pi$, $n\in\Z$:
\begin{align}\label{LCT}
\psi_x^\theta(t)&=\frac{1}{\sqrt{2\pi|\sin\theta|}}\exp\left(-i\frac{1}{2\tan\theta}t^2+i\frac{x}{\sin\theta}t\right).
\end{align}
For the case $\theta=n\pi$, the vectors are the Dirac distributions:
$$\psi_x^{n\pi}(t)=\left\{\begin{array}{ll}\delta(t-x)&\text{if $n$ is even or 0}\\\delta(t+x)&\text{if $n$ is odd}\end{array}\right..
$$
The parameter $\theta$ allows to pass from the time representation of the signal ($\theta=0$) to the frequency one ($\theta=\pi/2$) via intermediate representations. In fact, each $\psi_x^{\theta}$ is a linear chirp with phase derivative: 
$\phi'(t)=-\frac{1}{\tan\theta} \cdot t+ \frac{x}{\sin\theta}$.
Thus, this transform can be seen as the projection on a basis of linear chirps.
%%%%%%%%%%%%%%%%
\subsubsection*{Properties and relations with other transforms}
The linear chirp transformation is closely related to the theoretical work of~\cite{MV,MMV} in quantum mechanics where the tomogram transform is introduced. Several selfadjoint operators with their eigenvectors are presented and among them the operator $B(\theta)$. The tomogram transform associated to $B(\theta)$ consists in the analysis of the energy density of the signal when projected onto the eigenvectors of this operator. The LCT is connected to the selfadjoint operator $A_L(\theta)$:
\begin{align}\label{B}
A_L(\theta)=\cos\theta \cdot t-i\sin\theta\cdot \frac{d}{dt}=B(-\theta),
\end{align}
where $t$ and $d/dt$ are respectively the $t$-multiplication operator and the derivative operator. It acts on the Hilbert space $L^2(\R)$ and its generalized eigenfunctions, are the set of linear chirps $\{\psi_x^{\theta}\}_x$. That is to say: $A_L(\theta)\psi_x^{\theta}=x\psi_x^{\theta}$. Although coming from the same operator, the LCT and the Tomogram transform are not designed for the same purpose. Whereas the tomogram is designed to study the distribution of energy in a signal, the LCT stresses more on the extraction and separation of different parts of a signal, in a fast numerical manner.
%The tomogram transform provides interesting results concerning its application to the study of plasma reflectometry signals~\cite{BLMV1,BLV2}. We expand them in section~\ref{experiment} by means of the LCT.

The goal of this subsection is to show two important properties concerning the use and computation of the LCT, in particular its possibility to be implemented with a fast algorithm. It is deduced from the operator approach. For $\theta\neq n\pi$, since
$$\frac{d}{dt}\left(\exp(i\frac{1}{2\tan\theta}t^2)f(t)\right)=\exp(i\frac{1}{2\tan\theta}t^2)\left(\frac{i}{\tan\theta}t+\frac{d}{dt}\right)f(t),
$$
the operator $A_L(\theta)$ can be written equivalently as:
\begin{equation}\label{AL1}
A_L(\theta)=\exp(-i\frac{1}{2\tan\theta}t^2)\left(-i\sin\theta\frac{d}{dt}\right)\exp(i\frac{1}{2\tan\theta}t^2).
\end{equation}
Moreover, let $D_{\theta}$ be the dilatation operator such that:
$$(D_{\theta}f)(t)=\sqrt{|\sin\theta|}f(\sin\theta t).
$$
It is unitary and its inverse is
$$(D_{\theta}^{-1}f)(t)=\frac{1}{\sqrt{|\sin\theta|}}f(\frac{t}{\sin\theta}).
$$
Since for any $f$
$$D_{\theta}^{-1}\frac{d}{dt}D_{\theta}f=\sin\theta \frac{d}{dt}f,
$$
the operator $A_L(\theta)$ can be written as
\begin{align}\label{fB}
A_L(\theta)&=\exp(-i\frac{1}{2\tan\theta}t^2)D_{\theta}^{-1}\left(-i\frac{d}{dt}\right)D_{\theta}\exp(i\frac{1}{2\tan\theta}t^2)\\
&=\exp(-i\frac{1}{2\tan\theta}t^2)D_{\theta}^{-1}A_L(\frac{\pi}{2})D_{\theta}\exp(i\frac{1}{2\tan\theta}t^2)\nonumber.
\end{align}
Eq.~\eqref{fB} shows that for all $\theta\neq n\pi$:
\begin{itemize}
 \item[i)] since $D_{\theta}^{-1}$ and $\exp(it^2/2\tan\theta)$ are analytic with respect to $\theta$, the eigenvectors are analytic with respect to $\theta$ (see chap. 7 of~\cite{K}) and so are the chirp transform components $C_L^{\theta}$. This is an important property: the smoothness with respect to $\theta$ guarantee the robustness of the method i.e. a small change in $\theta$ causes a small change of $C_L^{\theta}$.
\item[ii)] $A_L(\theta)$ is unitary equivalent to the derivative operator. Then the eigenvectors $\{\psi_x^\theta\}_x$ of~\eqref{B} are related to the eigenvectors $\{\psi_x^{\pi/2}\}_x$ of the derivative operator i.e. the Fourier basis $\{e^{ixt}\}_x$.  Rel.~\eqref{fB} implies:
\begin{align}\label{psiLi}
\psi_x^{\theta}(t)&=e^{-i\frac{1}{2\tan\theta}t^2}D_{\theta}^{-1}\psi_x^{\pi/2}(t).
\end{align}
Hence the projection can be computed using the Fourier transform as ($D_{\theta}$ is unitary so that $(D_{\theta}^{-1})^*=D_{\theta}$):
$$(f,\psi_x^\theta)=(f,e^{-i\frac{1}{2\tan\theta}t^2}D_{\theta}^{-1}\psi_x^{\pi/2})=(D_{\theta}e^{i\frac{1}{2\tan\theta}t^2}f,\psi_x^{\pi/2}).
$$
As a consequence, a fast Fourier transform algorithm can be used to compute the chirp transform.
The action of $D_{\theta}e^{i\frac{1}{2\tan\theta}t^2}$ can be seen as a deformation of the time-frequency plane proportional to $\theta$. Linear chirps become stationary signals and they can be decomposed efficiently by a Fourier transform. From this point of view, the LCT transformation may be seen as a two steps process: first a smooth deformation of the function and second a projection onto the Fourier basis.
\end{itemize}
Eventually, it is worth noticing that the linear chirp transform is related to the fractional Fourier transform $X_{\alpha}(u)$ as defined in~\cite{A94} by:
$$X_{\alpha}(u)=\sqrt{\frac{1-i\cos\alpha}{\sin\alpha}}\exp(\frac{u^2}{2\tan\alpha})(C_L^{\alpha}f)(u).
$$
This latter transform is connected to fractional calculus and fractional derivatives~\cite{BM}.
%%%%%%%%%%%%%%%%%%
\subsubsection*{Definition on the interval}
For signals of finite time length $T$, and for $\theta\ne n\pi$, $n\in\Z$, the $\psi_x^{\theta}$ on the domain $[0,T]$ have the following expression~\cite{BLMV1}:
\begin{align}\label{psiL}
\psi_x^{\theta}(t)=\frac{1}{\sqrt{T}}\exp\left(-i\frac{1}{2\tan\theta}t^2+i\frac{x}{\sin\theta}t\right).
\end{align}
They form an orthonormal basis of the space $L^2(0,T)$. The real quantity $x$ is the associated eigenvalue which takes discrete values in this case:
\begin{align}
x=\frac{2\pi\sin\theta}{T}m,\quad m\in\Z.
\end{align}
In order to keep the domain bounds unchanged, the relation given in Eq.~\eqref{fB} is modified to:
\begin{align}
A_L(\theta)=\exp(-i\frac{1}{2\tan\theta}t^2)\widetilde{D}_{\theta}^{-1}A_L(\frac{\pi}{2})\widetilde{D}_{\theta}\exp(i\frac{1}{2\tan\theta}t^2),
\end{align}
where
$$\widetilde{D}_{\theta}^{-1}\psi_x^{\pi/2}(t)=\psi_{x/\sin\theta}^{\pi/2}(t).
$$
Then the previous properties $i)$ and $ii)$ hold and:
\begin{align}\label{LCTFFT}
(f,e^{-i\frac{1}{2\tan\theta}t^2}\widetilde{D}_{\theta}^{-1}\psi_x^{\pi/2})=&\frac{1}{\sqrt{T}}\int_{0}^{T}e^{i\frac{1}{2\tan\theta}t^2}f(t)\overline{\psi_{x/\sin\theta}^{\pi/2}(t)}dt\\
=&(e^{i\frac{1}{2\tan\theta}t^2}f,\psi_{x/\sin\theta}^{\pi/2})\nonumber.
\end{align}
This insures that, even in the finite time context, a fast algorithm can be used.

Eventually, let us state a remark on the boundary conditions of the domain. A function possessing a discontinuity involves a Fourier series containing a large number of non negligible harmonics. For continuous signals satisfying $f(0)=f(T)$, the Fourier transform is well adapted and gives a compact expression.
This is also true for the linear chirp transformation if the following condition (given by Eq.~\eqref{LCTFFT}) is satisfied:
\begin{equation}\label{LCTbound}
f(0)=e^{i\frac{1}{2\tan\theta}T^2}f(T).
\end{equation}
The parameter $\theta$ allows to tune the transformation in order to obtain this condition on a given signal $f$.
%%%%%%%%%%%%%%%%%%%

%%%%%%%%%%%%%%%%%%%%%%%%%%%%%%%%%%%%%%%%%
\subsection{The Hyperbolic Chirp Transformation}
%%%%
\subsubsection*{Definition}
This transformation is defined on $L^2(\R_+)$. Let $c$ be a real valued function of $t\in\R_+$. The vectors of the basis associated to the hyperbolic chirp transformation are the following ones:
\begin{equation}\label{HCT}
\varphi_x^{\theta}(t)=\frac{1}{\sqrt{2\pi|\sin\theta|t}}\exp\left(-i\frac{1}{\tan\theta}c(t)+i\frac{x}{\sin\theta}\ln t\right).
\end{equation}
Let $f\in L^2(\R^+)$. For all $\theta\neq n\pi$, $n\in\Z$, we define the hyperbolic chirp transform $C_{H,c}^{\theta}$ of $f$ by:
\begin{equation}
C_{H,c}^{\theta}(x)=(f,\varphi_x^{\theta}).
\end{equation}
For the case $\theta=n\pi$, the vectors are the Dirac distributions:
$$\varphi_x^{n\pi}(t)=\left\{\begin{array}{ll}\delta(t-x)&\text{if $n$ is even or 0}\\\delta(t+x)&\text{if $n$ is odd}\end{array}\right..
$$
The parameter $\theta$ allows to pass from the time representation of the signal ($\theta=0$) to the ``scale representation'' ($\theta=\pi/2$) via intermediate ones. For $\theta=\pi/2$, the hyperbolic chirp transform is related to the operator generator of dilatations.
For $\theta\neq n\pi$, each $\psi_x^{\theta}$ can be seen as a hyperbolic chirp with phase derivative: 
$\phi'(t)=-\frac{1}{\tan\theta}c'(t) + \frac{x}{\sin\theta}\frac{1}{t}$.

%%%%%%%%%%%%%%%%
\subsubsection*{Properties and relations with other transforms}
The Hyperbolic Chirp Transformation can be related to other transformations for particular values of $c$ and $\theta$. For example it is a generalization of the second tomogram transform suggested in~\cite{MV,MMV}: the time-scale tomogram is the HCT with $c(t)=t$.

Let us show how the HCT can be implemented with a fast algorithm.
Let $D$ be the generator of dilatations acting in $L^2({\R_+})$ and defined by:
$$D=\frac{i}{2}(t\frac{d}{dt}+\frac{d}{dt}t)=i(t\frac{d}{dt}+\frac{1}{2}).
$$ 
The HCT is associated with the following operator:
\begin{align}\label{B2}
A_H(\theta)=\cos\theta\cdot t\ c'(t)-\sin\theta \cdot D=\cos\theta\cdot t\ c'(t)-i\sin\theta\cdot (t\frac{d}{dt}+\frac{1}{2}),
\end{align}
where $c'$ stands for the derivative of $c$.
Indeed, for $\theta\neq n\pi$, $\varphi_x^{\theta}$ are the eigenvectors of $A_H(\theta)$:
\begin{align*}
\frac{d\varphi_x^{\theta}}{dt}(t)=\left(i\frac{x}{t\sin\theta}-i \frac{c'(t)}{\tan\theta}-\frac{1}{2t}\right)\varphi_x^{\theta}(t)&\Leftrightarrow\\
it\sin\theta\frac{d\varphi_x^{\theta}}{dt}(t)=\left(-x+tc'(t)\cos\theta-i\frac{\sin\theta}{2}\right)\varphi_x^{\theta}(t)&\Leftrightarrow\\
tc'(t)\cos\theta\varphi_x^{\theta}(t)-i\sin\theta (t\frac{d}{dt}+\frac{1}{2})\varphi_x^{\theta}(t)=x\varphi_x^{\theta}(t)&.
\end{align*}
Since $D$ is selfadjoint and $c'$ is a real-valued function, the operator $A_H(\theta)$ is selfadjoint and as a consequence (spectral theorem) the vectors $\{\varphi_x^{\theta}\}_x$ are orthonormal and~\eqref{resynt} holds.
Similarly to the LCT case, the operator $A_H(\theta)$ can be put under the form:
\begin{equation}\label{BthetaD}
A_H(\theta)=-\sin\theta e^{-i\frac{c(t)}{\tan\theta}}De^{i\frac{c(t)}{\tan\theta}}.
\end{equation}
When $\theta=\pi/2$, $A_H(\pi/2)=-D$ and the HCT is (up to a normalization constant) the Mellin transform (as defined in~\cite{Ob}) where the real part of the exponent is restricted to the value $1/2$ and the imaginary part is equal to $(-x)$: 
\begin{equation}
C_{H,c}^{\pi/2}(x)=\int_0^{\infty}\frac{1}{\sqrt{2\pi t}}\exp\left(-ix\ln t\right)f(t)dt=\frac{1}{\sqrt{2\pi}}\int_0^{\infty} t^{(-ix+\frac{1}{2})-1}f(t)dt.
\end{equation}
%Let us note that the above formula defines a unitary mapping between $L^2(\R_+)$ and $L^2(\R)$. 
Then the HCT can be written in term of Mellin transform since from Eq.~\eqref{BthetaD} the following relation can be deduced:
\begin{equation}
\varphi_x^{\theta}= e^{-i\frac{c(t)}{\tan\theta}}\varphi_{x/\sin\theta}^{\pi/2}\qquad\text{and}\qquad (f,\varphi_x^{\theta})=(e^{i\frac{c(t)}{\tan\theta}}f,\varphi_{x/\sin\theta}^{\pi/2}).
\end{equation}
In conclusion, let us remark that fast algorithms exist for the computation of the Mellin Transform~\cite{BBO}.

%The case where $c$ is a complex-valued function can be considered. Under this assumption the operator associated is not anymore selfadjoint and its eigenvectors are not orthogonal. Nevertheless they can be calculated and are bi-orthogonal so that there exists a formula similar to~\eqref{resynt} to reconstruct the signal. If $c(t)=i(\ln t)^2$ for example, Eq.~\eqref{HCT} is the Altes wavelet~\cite{Altes}. This function is the logarithmically warped version of a Gaussian. It minimizes the time-scale uncertainty as the Gaussian law minimizes the time-frequency uncertainty.
%%%%
\subsection{Self-similar and stationary processes}
A one-to-one connection has been established between stationary and self-similar stochastic processes in~\cite{FBA}. It is done by means of the Lamperti transformation. We show that the same process allows to connect the LCT and the HCT together. Since the LCT is efficient for detecting deviations from stationarity, this demonstrates that the HCT is appropriate to detect deviation from self-similarity. 

Let us briefly recall the Lamperti transformation. Let $f$ be a complex-valued function of $t\in\R$. The Lamperti transform of $f$ is defined for $a>0$ and $t>0$ by:
\begin{equation}
(L_af)(t)=t^af(\ln t).
\end{equation}
It has for inverse transform, for $g:\R_+\to\C$:
\begin{equation}
(L_a^{-1}g)(t)=e^{-at}g(e^t),\ t\in\R.
\end{equation}
To show the relation between LCT and HCT we exhibit the link between their respective associated operator. For this purpose, we need to extend the definition of~\cite{FBA} to $a=-1/2$. Since
\begin{align*}
&(L_{-1/2}f,L_{-1/2}f)_{L^2(\R_+)}=\int_{0}^{\infty}t^{-1}|f(\ln t)|^2dt=\int_{-\infty}^{\infty}|f(\tau)|^2d\tau=(f,f)_{L^2(\R)},\\
&(L_{-1/2}^{-1}f,L_{-1/2}^{-1}f)_{L^2(\R)}=\int_{\R}e^{t}|f(e^t)|^2dt=\int_{0}^{\infty}|f(\tau)|^2d\tau=(f,f)_{L^2(\R_+)},
\end{align*}
$L_{-1/2}$ is well defined as a unitary operator from $L^2(\R)$ to $L^2(\R_+)$.
The last step of the proof consists in noticing the following two relations: first we have a correspondence between the derivative operator and the dilatation operator,
\begin{align*}
L_ai\frac{d}{dt}L_a^{-1}g(t)=&L_a\left(-aie^{-at}g(e^t)+ie^{-at}e^tg'(e^t)\right)\\
=&-ag(t)+tg'(t)\\
=&-i(a+\frac{1}{2})g(t)+Dg(t).
\end{align*}
Secondly, the Lamperti operator turns a multiplication operator by a complex-valued function $h$ into another multiplication operator:
\begin{align}
 L_ahL_a^{-1}(t)=h(\ln t).
\end{align}
As a consequence one has:
\begin{align}\label{LHT}
L_a(\cos\theta t-i\sin\theta \frac{d}{dt})L_a^{-1}=\cos\theta \ln t-\sin\theta D+i\sin\theta (a+\frac{1}{2}).
\end{align}
Then  $L_{-1/2}$ makes the connection between the LCT and the HCT with $c(t)=\ln t$. More precisely, let $\{\widetilde{\varphi}_x^{\theta}\}_x$ be the set of eigenvectors of the operator $A_H(\theta)$ with $c(t)=\ln t$, then 
\begin{equation}
\psi_x^{\theta}=L_{-1/2}\widetilde{\varphi}_x^{\theta}\qquad\text{and}\qquad C_L^{\theta}(f)=C_{H,\ln t}^{\theta}(L_{-1/2}^{-1}f).
\end{equation}

%For $a>0$, the Lamperti transformation is not unitary\footnote{However, for each $a$ the Lamperti transformation is unitary from $L^2(\R)$ to $L^2(\R_+,t^{-2a-1}dt)$.} from $L^2(\R)$ to $L^2(\R_+)$ and as a result the right hand side operator of~\eqref{LHT} is not selfadjoint anymore. Nonetheless, the (unnormalized, non orthogonal) vectors can be calculated:
%\begin{align}
%\psi_a^{\theta}(t)=\exp\left(-i\frac{x+i\sin\theta(a+1/2)}{\sin\theta}\ln t-\frac{1}{2}\ln t+\frac{i}{2\tan\theta}(\ln t)^2\right).
%\end{align}

%%%%%%%%%%%%%%%%%%%%%%%%%%%%%%%
\section{Linear chirp transformation and turbulent plasma}\label{experiment}
%%%%%%%%%%%%%%%%%%%%%%%%%%%%%%%%%%%%%%%

A plasma of density $d$ has the property to absorb, reflect or be transparent to a wave, depending on its frequency. The density of the reflective layer can then be deduced from the frequency of the incident wave. Reflectometry consists in emitting a wave during 20 $\mu s$ with a frequency increasing linearly with time, and detecting the reflected echo. The phase difference between emitted and reflected waves gives the position of the plasma reflective layer. Thus a well chosen sweep of frequencies allows to probe the density in depth and obtain the density profile. It can be reconstructed from the output signal of the reflectometer through a mathematical relation~\cite{BCI87}.
The calculation of the density profile from this relation is based on the assumption that the signal is of the form:
$$y(t) = A(t) e^{i\phi(t)},$$
with a slowly varying amplitude $A$ compared to the variations of $e^{i\phi}$. Furthermore, at each time $t$ a single frequency $F_b(t)$ (beat frequency) must be associated to $\phi(t)$ via the relation:
$$F_b=\frac{1}{2\pi}\frac{d\phi}{dt}.
$$
This latter assumption fails when multi reflections occur as each echo adds its beat frequency to the phase $\phi$.
A part of them is due to reflections on the walls surrounding the plasma while the other part is caused by abrupt density fluctuations, resulting from the plasma turbulence. The former must be filtered out whereas the latter must be taken into account in the calculations since it is a consequence of the density shape. A previous work on the subject~\cite{BLMV1,BLV2} has shown the ability of the time-frequency tomogram to separate the plasma signal from the undesired wall echoes. As pointed out in the previous section, this tomogram is closely related to the linear chirp transform. However, it does not treat the plasma multi reflections and it is still not possible to obtain a correct density profile. The following analysis shows that the linear chirp transform can be applied to separate these multi reflections. This could lead to a more accurate reconstruction of the density profile and a better understanding of the plasma turbulent behavior.

\begin{figure}
\center
\begin{minipage}{0.5\linewidth}
\includegraphics[width=8cm]{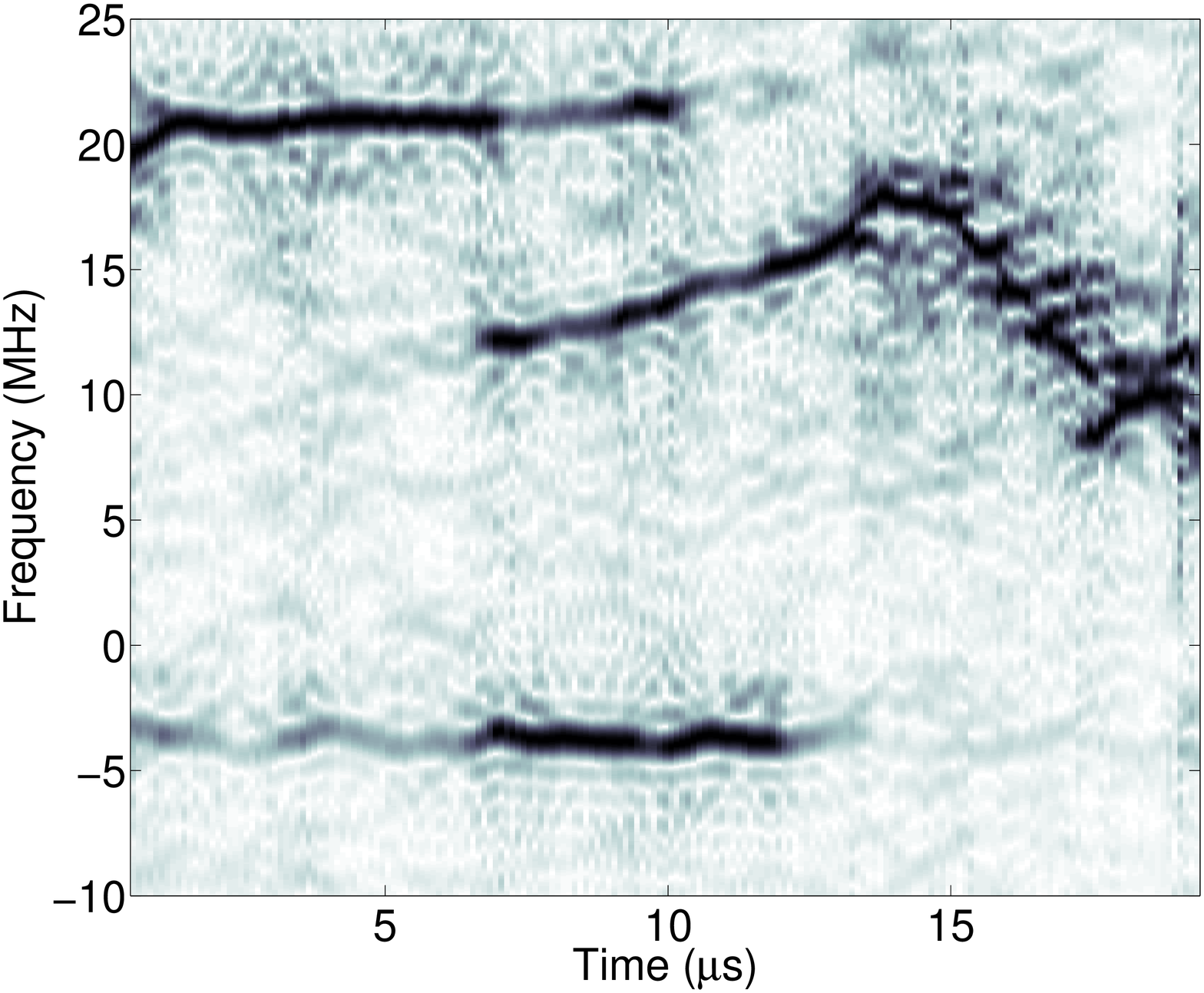}
\end{minipage}%
\begin{minipage}{0.5\linewidth}
\includegraphics[width=8cm]{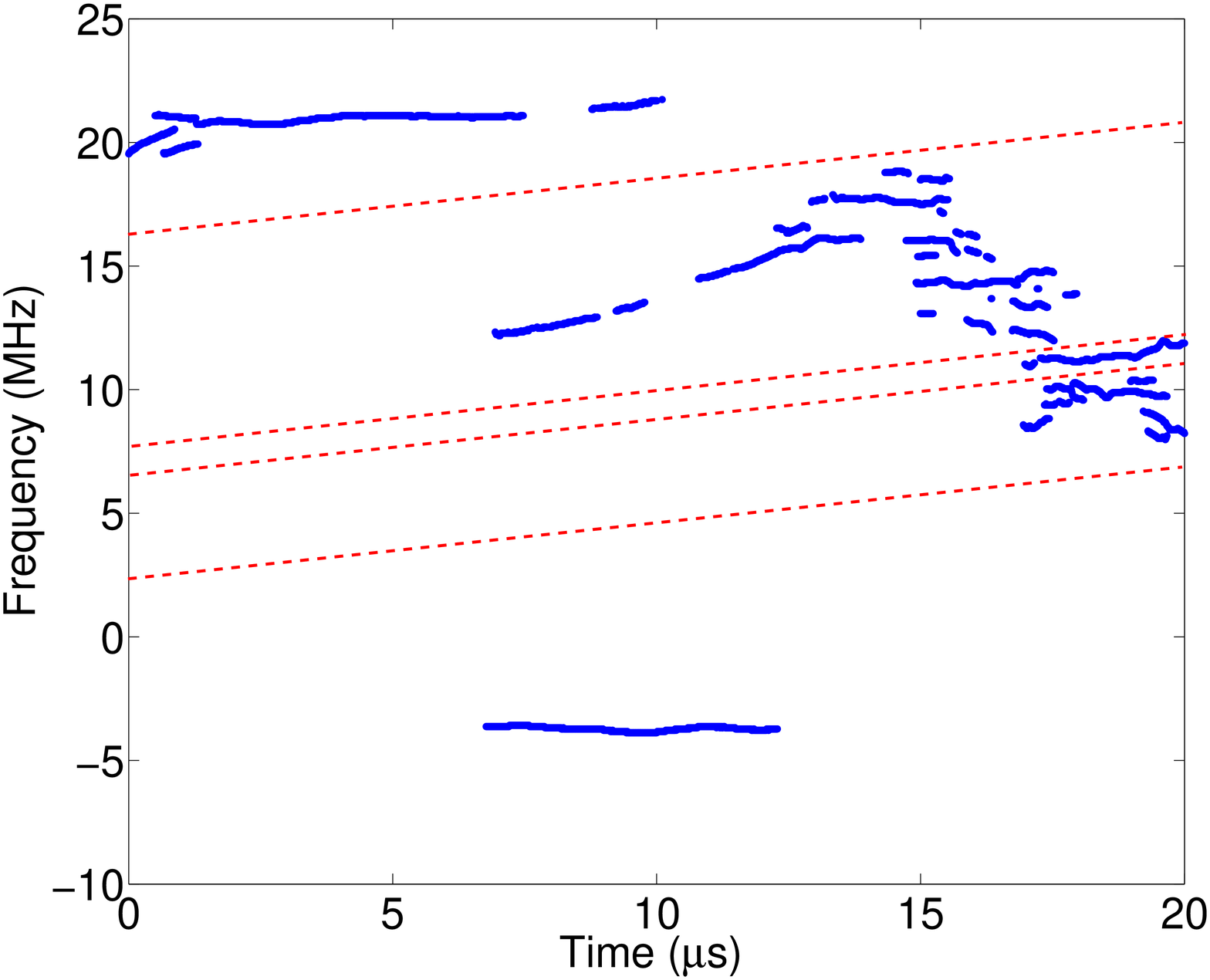}
\end{minipage}
\caption{(a) Spectrogram of the reflectometer output (Sliding FFT). (b) Ridges of the spectrogram. The linear chirp transform can cut the time-frequency plane with a specified angle $\theta$; red lines delimit the different parts separated by LCT for $\theta=\pi/5$.}\label{1003f2}
\end{figure}

The spectrogram of the reflectometer output signal has been plotted on Fig.~\ref{1003f2}-(a). The (multivalued) beat frequency can be seen in black.
At the beginning of the sweep the plasma is transparent for the wave: in that region, the reflection of the wave is due solely to the porthole and the back wall. Since these obstacles are fixed, the fight time of the reflected wave is constant and $y$ contains two components of constant frequency. As the porthole is close to the reflectometer, it gives a low beat frequency component (around $-5$ MHz) whereas the back wall is associated to a higher frequency one (around $20$ MHz). After a certain time, the frequency of the emitted wave reach the first plasma cutoff frequency and the plasma start to be reflective. The plasma reflection appears at 6 $\mu s$. In the same time, the back wall reflection intensity decreases and disappears since the plasma screens the wall. The beat frequency of the plasma-reflected wave increases from 10 to 20 MHz as the frequency of the emitted wave increases: the wave enters deeper and deeper inside the plasma. A change in the reflection behavior appears after 12 $\mu s$ where the beat frequency starts to decrease and is not anymore linear and single valued, revealing a turbulent zone in the plasma.

\begin{figure}
\center
\includegraphics[width=8cm]{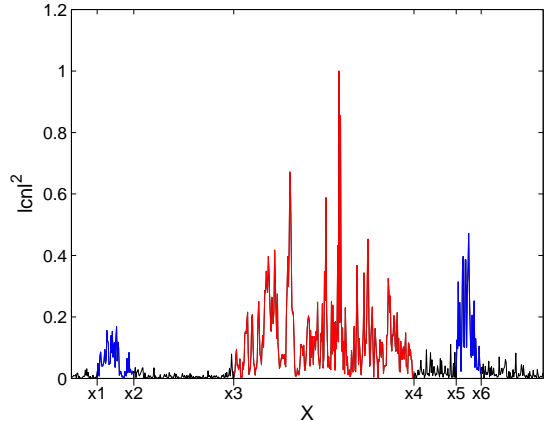}
\caption{Square modulus of the linear chirp transform $C_L^{\theta}$ of the reflectometry signal for $\theta=\pi/5$.}\label{1003f5}
\end{figure}

\begin{figure}
\center
\includegraphics[width=7cm]{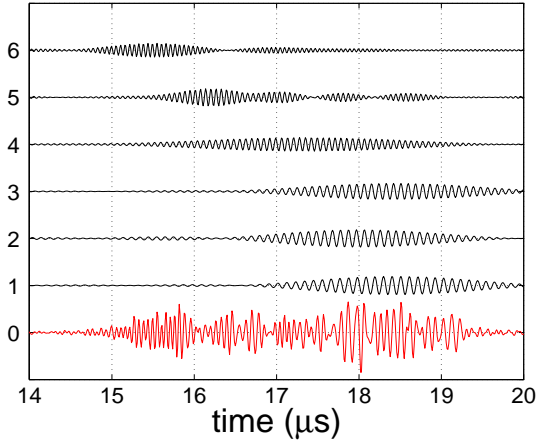}
\caption{
Re-synthesis of the separated plasma multi reflections by the linear chirp transformation. Each component (1 to 6) corresponds to a different ridge in the turbulent region of the time-frequency plane. The bottom one is the raw signal.}\label{figure9}
\end{figure}

In order to analyze the plasma multi reflections, the ridge detection algorithm of~\cite{Chandre} has been applied to the signal. The result is plotted on Fig.~\ref{1003f2}-(b). The chaotic region appears to be a superposition of small chirps where many of them have a similar slope. This is exactly the type of pattern which can be separated by the linear chirp transform. The dashed red lines on the graph are chirps of the LCT with a parameter $\theta=\pi/5$. They are the delimiter of different domains of the time-frequency plane. Two of them separate the porthole and back wall from the plasma reflection (the ones with a starting frequency of 2.5 and 16 MHz respectively). In addition, the plasma multi reflections can be isolated: an example is given where two chirps starting at 6 and 7 MHz frame a single component of the multi reflections. The LCT of the reflectometry signal has been performed for $\theta=\pi/5$ and the square modulus is plotted on figure~\ref{1003f5}.  Up to a constant ($\sin(\pi/5)$), the eigenvalue $x$ corresponds in practice to the initial frequency of the chirp (at $t=0$) used for the projection. Let us recall that the slope of the chirp is given by $1/\tan(\pi/5)$. The square modulus of $C_L^{\pi/5}$ corresponds to the amount of energy contained in the signal projection.
As expected, three regions of high energy can be seen distinctively on the graph. Defined by the intervals $[X_1,X_2]$, $[X_3, X_4]$ and $[X_5, X_6]$ are the porthole, plasma and back wall regions respectively. The middle zone corresponds to the plasma reflection with its multi reflections: each peak and its direct neighborhood corresponds to a particular ridge of the time-frequency plane. If the ridge is sharp and possesses a slope close to $1/\tan(\pi/5)$ the peak is narrow, otherwise it broadens as a part of the ridge energy is shared between several nearby chirps. By setting a threshold value on $|C_L^{\theta}|$, several peaks can be isolated. Then the associated time signals are synthesized by inverse transform following~\eqref{resynt}. The sum in this formula is taken over $x$ in the respective peak intervals. These signals are shown on Fig.~\ref{figure9}. Each one corresponds to a ridge of the time-frequency plane. The bottom one is the sum of the multi reflections.

%%%%%%%%%%%%%%%%%%%%%%%%%%%%%%%%%
\section{Conclusion}\label{conclusion}

We presented a method for the analysis of chaotic signal involving two new transformations. It is based on the extraction of ridges in the time-frequency domain. 
%Combining the linear chirp transform with the ridge detection algorithm of~\cite{Chandre}, suggests a way to extract the chaotic parts of a signal. 
The ability to isolate the chaotic behavior (or a part of it) from a raw signal could considerably ease the interpretation of nonlinear effects. The results obtained in the case of plasma turbulences are promising.
The free parameter $\theta$ in the linear chirp transformation gives a flexibility which allows to adapt the separation to the shape of the ridges in the time-frequency plane. This is much more powerful than a Fourier transform, a frequency filtering or a wavelet transform. In addition, this choice is robust, i.e. a small error in the estimation of the best $\theta$ will not affect dramatically the result. This property also apply to the HCT when analyzing self-similar signals. The function $c$ in the HCT gives an additional degree of freedom and open a way toward improved chaos detection methods. 

One of the major results concerns the algorithms of the transformations. The chirp transformations and their inverse are done using a relatively simple implementation, thanks to the orthogonality property. Moreover, fast algorithms (fast Fourier transform) can be used for the computation. This is an important property if the computation of many decomposition is needed or if the result is demanded immediately for a fast reaction in a control purpose.

Concerning the relation established between the LCT, the HCT and the Lamperti transform, it shows that the HCT could be a tool well suited for the detection of chaos in self-similar processes. In a analogous manner as what the LCT analysis obtains for square integrable signals, it could extract chaotic regions in a signal issued from stochastic motions. Thus the present study shows a way to extend the ridge analysis to theses types of signals.

\section*{Acknowledgements}
The authors wish to thank X. Leoncini for useful remarks and suggestions, C. Chandre for his help with the ridge detection algorithm and F. Clairet from the Tore Supra team (Commissariat \`a l'\'energie atomique, Cadarache, France) for providing us the reflectometry data.

%%%%%%%%%%%%%%%%%%%%%%%%%%%%%%%%%

\end{document}